\documentclass[11pt]{article}

\input epsf
\usepackage{latexsym}
\usepackage{amssymb}
\usepackage{amsmath}
\usepackage[dvips]{graphicx}

\begin{document}
\centerline{\Large \bf Asymptotic behaviour of curvature 
and matter in the
Penrose limit}
\vskip 2cm

\centerline{Kerstin E. Kunze\footnote{E-mail: Kerstin.Kunze@cern.ch}}

\vskip 0.3cm

\centerline{{\sl F\'\i sica Te\'orica,  
Universidad de Salamanca,}}
\centerline{{\sl Plaza de la Merced s/n,
37008 Salamanca, Spain.}}

\vskip 1.5cm

\centerline{\bf Abstract}

\vskip 0.5cm

\noindent
The asymptotic behaviour of the components of the Weyl tensor
and of the energy-momentum tensor
in the Penrose limit is determined. 
In both cases a peeling-off property is found.
Examples of different types of matter 
are provided.
The  expansion and shear of 
the congruence of null geodesics 
along which the Penrose limit is taken 
are determined. Finally, the
approach to the singularity in the Penrose limit 
of cosmological space-times is discussed.

\vskip 1 cm

\section{Introduction}

Recent developments in string-/M theory 
have let to renewed interest in an argument by Penrose 
\cite{penrose} that in a neighbourhood of a null geodesic,
which contains no conjugate points,
any space-time has a plane wave as a limit 
(\cite{blau} and references therein). 
The Penrose limit construction involves changing first
to a suitable system of coordinates and then
rescaling by a constant parameter $\Omega$ the coordinates
as well as the metric. In the limit $\Omega\rightarrow 0$ a plane wave
space-time is obtained. The parameter $\Omega$
can be interpreted as an additional coordinate.
Thus lifting an $n$-dimensional space-time to an $(n+1)$-
dimensional space-time with boundary, 
whose boundary at $\Omega=0$ is given by
the Penrose limit \cite{penrose}.

The Penrose limit construction allows to associate any
space-time with a plane wave space-time.
Another situation in which a space-time approaches a 
plane wave is in the case of a metric far away from
an isolated radiating source \cite{Bondi}. In this case the space-time
approaches globally a plane wave whereas in 
the case of the Penrose limit the plane wave is local in the 
neighbourhood of a particular geodesic.

In the case of the limiting space-time far away 
from an isolated radiating source the approach to 
the limiting plane wave space-time has been 
determined. Namely, the  
components of the Weyl tensor show a typical
peeling-off behaviour \cite{Bondi}. In the Newman-Penrose
formalism the components of the Weyl tensor are 
described by five complex scalars $\Psi_i$. 
In this particular case, as the far field limit is approached 
these five components become negligible one after 
the other until only $\Psi_4$ remains which 
is associated with a plane wave space-time.
A peeling-off behaviour was also found  
in space-times admitting two space-like Killing 
vectors \cite{sp-K}.  Furthermore,
space-times describing light like
signals propagating through a general Bondi-Sachs
space-time have peeling properties as well. In 
certain cases these are different from those
of a Bondi-Sachs space-time \cite{bh}.

Here the asymptotic approach in the Penrose
limit to a plane wave will be investigated.
Locally any space-time approaches a plane 
wave along a null geodesic. Thus this is 
a general feature of any metric.
Similarly to the global plane wave limit of 
the isolated radiating source space-times and those
admitting two space-like Killing vectors,
a peeling-off behaviour of the
curvature as well as the components of
the energy-momentum tensor is found in the 
Penrose limit.

Plane wave space-times are exact classical
string vacua \cite{ak}. 
Space-times in general are no such solutions,
even though vacuum solutions of general relativity 
are always solutions to string theory in the
low energy limit. 
Thus the approach to the Penrose limit 
might give  corrections to  exact classical string
vacua.

In section 2, the Penrose limit construction 
is formulated using the Newman-Penrose formalism which
is especially adapted to the treatment of 
null geodesics. Then the peeling-off property 
of the Weyl  tensor and of the
energy-momentum tensor is found. In 
section 3 the kinematics of the null geodesics is
determined and the approach to  null
power law singularities is discussed.
Finally, in section 4 conclusions are presented.

\section{The Weyl tensor and energy-momentum tensor in the Penrose limit}

Following  \cite{penrose} the metric in the neighbourhood of a
segment of a null geodesic containing no conjugate points can be
written as
\begin{eqnarray}
ds^2=dudv+\alpha dv^2+\beta_1 dxdv+\beta_2 dydv-\frac{2e^{-\mu_1}}{Z+
\bar{Z}}\left(dx+iZdy\right)\left(dx-i\bar{Z}dy\right),
\label{pen-met}
\end{eqnarray}
where $\alpha, \beta_i, Z$ and $\mu_1$ are functions of all
coordinates $u,v,x,y$. $Z$ is a complex function. Complex 
conjugation is denoted by a bar.
The parametrization of the 
components of the metric over the two-dimensional
$x-y$ subspace is of the form used in colliding plane wave
space-times \cite{gri,kk}. Whereas in the case of collding plane waves
$Z$ is just a function of $u$ and $v$, here $Z$ is a function of
all variables. This form of the metric seems to be 
convenient, since plane waves, and thus the resulting metrics in
the Penrose limit, are a particular solution of 
colliding plane wave space-times.

Recently, the Penrose limit has been formulated in a 
covariant form by making use of a parallel-transported
frame along the chosen null geodesic \cite{cov-pen1,cov-pen2,pap}.
Such a frame is obtained by introducing
a null tetrad with two real null vectors 
$l^{\mu}$, $n^{\mu}$ and
two complex null vectors which are complex conjugates of each others
$m^{\mu}$, $\bar{m}^{\mu}$. They satisfy the relations \cite{chandra,gri}
\begin{eqnarray}
l_{\mu}n^{\mu}=1,\hspace{2cm} m_{\mu}\bar{m}^{\mu}=-1,
\hspace{2cm}
g_{\mu\nu}=l_{\mu}n_{\nu}+n_{\mu}l_{\nu}-m_{\mu}\bar{m}_{\nu}
-\bar{m}_{\mu}m_{\nu}.
\end{eqnarray}

Explicit expressions for the null tetrad and 
the spin coefficients in the 
Newman-Penrose formalism are given in the appendix.

The Penrose limit  uses the freedom of rescaling coordinates
by a constant parameter. Namely, new coordinates $U,V,X,Y$ are 
introduced as follows \cite{penrose}
\begin{eqnarray}
u=U, \hspace{1cm} v=\Omega^2 V, \hspace{1cm} x=\Omega X, \hspace{1cm} 
y=\Omega Y,
\label{p1}
\end{eqnarray}
where $\Omega>0$ is a constant parameter.
Furthermore, a conformal transformation is applied to the metric
(\ref{pen-met})
such that 
\begin{eqnarray}
ds^2_{\Omega}=\Omega^{-2}ds^2,
\label{o-met}
\end{eqnarray}
resulting in 
\begin{eqnarray}
ds_{\Omega}^2=dUdV+\Omega^2 A dV^2+\Omega B_1 dXdV
+\Omega B_2 dY dV-\frac{2e^{-M}}{\zeta+\bar{\zeta}}
\left(dX+i\zeta dY\right)\left(dX-i\bar{\zeta} dY\right),
\label{e1}
\end{eqnarray}
where $A$, $B_i$ and $\zeta$ are functions of $U,V,X,Y$
and replace $\alpha$, $\beta_i$ and $Z$, respectively, of the 
metric (\ref{pen-met}).

The Penrose limit is defined as
\begin{eqnarray}
\hat{ds}^2={\rm lim}_{\Omega\rightarrow 0}ds^2_{\Omega}.
\label{p4}
\end{eqnarray}
Applying this to the metric (\ref{e1}) a plane wave
metric is obtained.
The approach to this plane wave, in powers of $\Omega$, 
will be studied by means of the 
components of the Weyl tensor. Furthermore, for 
non-vacuum space-times the asymptotic behaviour of the 
components of the energy-momentum tensor will be discussed.

In the Newman-Penrose formalism the 
ten components of the Weyl tensor are encoded in 
five complex scalars $\Psi_i$ (cf appendix).
These involve the spin coefficients as well as
their directional derivatives.
The Ricci identities \cite{chandra}
together with the asymptotic behaviour of the 
spin coefficients determine the scaling 
with the Penrose parameter $\Omega$ of the 
components of the Weyl tensor.

Using the expressions for the spin coefficients as given in the
appendix (cf equation (\ref{sp-c})) for the rescaled metric
(\ref{e1}) leads to the following overall
scaling with the 
Penrose parameter $\Omega$, which determines the factor
$\Omega^n$ in front of the $U-$ dependent part,
\begin{eqnarray}
\begin{array}{cccccc}
\kappa(\Omega)={\cal O}(\Omega^3) &\hspace{1.4cm}
\sigma(\Omega)={\cal O}(\Omega^2) &\hspace{1.4cm}
\lambda(\Omega)={\cal O}(\Omega^0)& \hspace{1.4cm} \nu=0\\
\rho(\Omega)={\cal O}(\Omega^2) &\hspace{1.4cm}
\mu(\Omega)={\cal O}(\Omega^0)&\hspace{1.4cm}
\tau(\Omega)={\cal O}(\Omega) &\hspace{1.4cm}
\pi(\Omega)={\cal O}(\Omega) \\
\epsilon(\Omega)={\cal O}(\Omega^2) &\hspace{1.4cm}
\gamma(\Omega)={\cal O}(\Omega^0) &\hspace{1.4cm}
\alpha_{NP}(\Omega)={\cal O}(\Omega) &\hspace{1.4cm}
\beta(\Omega)={\cal O}(\Omega).
\end{array}
\end{eqnarray}
The scaling with $\Omega$ 
of the directional derivatives (cf appendix equation (\ref{dd}))
is found by going back to the original coordinates $u,v,x,y$ since
all metric functions are functions of these coordinates and taking
partial derivatives with respect to the new coordinates
$U,V,X,Y$ leads to derivatives with respect to the old coordinates 
with an additional factor of $\Omega$. For example,
\begin{eqnarray}
D(\Omega)=-\Omega^2\frac{\alpha}{2}\frac{\partial}{\partial u}
+\Omega^2\frac{\partial}{\partial v},
\nonumber
\end{eqnarray}
where $\alpha$ is, as before, a function of $u, v, x , y$.
Therefore, the directional derivatives scale as follows
\begin{eqnarray}
D(\Omega)={\cal O}(\Omega^2) \hspace{1.2cm} 
\Delta(\Omega)={\cal O}(\Omega^0) \hspace{1.2cm} 
\delta(\Omega)={\cal O}(\Omega) \hspace{1.2cm} 
\bar{\delta}(\Omega)={\cal O}(\Omega).
\end{eqnarray}

Using the following Ricci identities \cite{chandra}
and the expression for 
$\Psi_2$ derived from Ricci identities 
\begin{eqnarray}
D\sigma-\delta\kappa&=&\sigma\left(3\epsilon-\bar{\epsilon}
+\rho+\bar{\rho}\right)
+\kappa\left(\bar{\pi}-\tau-3\beta-\bar{\alpha}_{NP}\right)
+\Psi_0\nonumber\\
D\beta-\delta\epsilon&=&\sigma\left(\alpha_{NP}+\pi\right)
+\beta\left(\bar{\rho}-\bar{\epsilon}\right)
-\kappa\left(\mu+\gamma\right)
-\epsilon\left(\bar{\alpha_{NP}}-\bar{\pi}\right)
+\Psi_1\nonumber\\
\Psi_2&=&\frac{1}{3}\left[D\gamma-\Delta\epsilon -\delta\alpha_{NP}
+\bar{\delta}\beta+D\mu-\delta\pi
-\left(\alpha_{NP}+\pi\right)\left(\tau+\bar{\pi}-\bar{\alpha}+\beta\right)
\right.
\nonumber\\
&&\left.
-\beta\left(\bar{\tau}+\pi+\alpha-\bar{\beta}\right)
+\left(\mu+\gamma\right)\left(\epsilon+\bar{\epsilon}+\rho-\bar{\rho}\right)
\right.
\nonumber\\
&&\left.
+\epsilon\left(\gamma+\bar{\gamma}+\mu-\bar{\mu}\right)
+2\nu\kappa-2\sigma\lambda
\right]\nonumber\\
\Delta\alpha_{NP}-\bar{\delta}\gamma&=&\nu\left(\rho+\epsilon\right)
-\lambda\left(\tau+\beta\right)+\alpha_{NP}\left(\bar{\gamma}-\bar{\mu}\right)
+\gamma\left(\bar{\beta}-\bar{\tau}\right)
-\Psi_3\nonumber\\
\Delta\lambda-\bar{\delta}\nu&=&-\lambda\left(\mu+\bar{\mu}
+3\gamma-\bar{\gamma}\right)+\nu\left(3\alpha_{NP}+\bar{\beta}+\pi-\bar{\tau}\right)
-\Psi_4
\end{eqnarray}
shows that 
the components of the Weyl tensor scale as
\begin{eqnarray}
\Psi_i(\Omega)={\cal O}(\Omega^{4-i}),
\end{eqnarray}
where $i=0,..,4$.
Thus the components of the curvature show 
a peeling-off behaviour.
The approach to the plane wave is determined
by the constant parameter $\Omega$.
This is different from the global
peeling-off behaviour found, for example,
in space-times describing the far-field
of an isolated radiating source. In this 
case the peeling-off behaviour
is controlled by one of the coordinates.
Namely, the components of the 
Weyl tensor behave as \cite{Bondi}
\begin{eqnarray}
\Psi_k={\cal O}(r^{k-5}),
\end{eqnarray}
in the limit $r\rightarrow\infty$ and 
$r$ is a radial coordinate.
Thus the dependence on this 
coordinate remains whereas 
this is not the case of the Penrose
limit. Here, $\Psi_4$ is independent
of the parameter $\Omega$.
As was already observed in the 
covariant approach to
the Penrose limit as formulated in 
\cite{cov-pen1,cov-pen2} here the 
component of the Weyl tensor $\Psi_4$
is exactly the same as in the original
space-time (\ref{pen-met}).

Using Einstein's equations in the tetrad basis
\begin{eqnarray}
R_{(a)(b)}-\frac{1}{2}\eta_{(a)(b)}R=-T_{(a)(b)}
\end{eqnarray}
allows to determine the asymptotic behaviour of the components
of the energy-momentum tensor.
In the Newman-Penrose formalism the components of the Ricci
tensor are denoted by $\Phi_{ab}$ (cf. appendix equation (\ref{ri})).
The Ricci identities as given in \cite{chandra}, e.g., 
can be used to find the scaling of $\Phi_{ab}$ and thus 
the behaviour of the components of the energy-momentum tensor.

The useful Ricci identities to determine the scaling of the 
$\Phi_{ab}$ are given by \cite{chandra}
\begin{eqnarray}
D\rho-\bar{\delta}\kappa&=&\left(\rho^2+\sigma\bar{\sigma}\right)
+\rho\left(\epsilon+\bar{\epsilon}\right)-\bar{\kappa}\tau
-\kappa\left(3\alpha_{NP}+\bar{\beta}-\pi\right)+\Phi_{00}
\nonumber\\
D\alpha_{NP}-\bar{\delta}\epsilon&=&
\alpha_{NP}\left(\rho+\bar{\epsilon}-2\epsilon\right)
+\beta\bar{\sigma}-\bar{\beta}\epsilon-\kappa\lambda-\bar{\kappa}\gamma
+\pi\left(\epsilon+\rho\right)+\Phi_{10}\nonumber\\
D\lambda-\bar{\delta}\pi&=&\rho\lambda+\bar{\sigma}\mu
+\pi\left(\pi+\alpha_{NP}-\bar{\beta}\right)
-\nu\bar{\kappa}-\lambda\left(3\epsilon-\bar{\epsilon}\right)+\Phi_{20}\nonumber\\
\delta\nu-\Delta\mu&=&\mu^2+\lambda\bar{\lambda}+\mu\left(\gamma+\bar{\gamma}\right)
-\bar{\nu}\pi+\nu\left(\tau-3\beta-\bar{\alpha}_{NP}\right)+\Phi_{22}\nonumber\\
\delta\gamma-\Delta\beta&=&\gamma\left(\tau-\bar{\alpha}_{NP}-\beta\right)
+\mu\tau-\sigma\nu-\epsilon\bar{\nu}-\beta\left(\gamma-\bar{\gamma}-\mu\right)
+\alpha\bar{\lambda}+\Phi_{12}\nonumber\\
\Delta\rho-\bar{\delta}\tau&=&-\left(\rho\bar{\mu}+\sigma\lambda\right)
+\tau\left(\bar{\beta}-\alpha_{NP}-\bar{\tau}\right)+\rho\left(\gamma+\bar{\gamma}\right)
+\nu\kappa-\Psi_2-2\Lambda\nonumber\\
\delta\alpha_{NP}-\bar{\delta}\beta&=&\mu\rho-\lambda\sigma
+\alpha_{NP}\bar{\alpha}_{NP}
+\beta\bar{\beta}-2\alpha\beta+\gamma\left(\rho-\bar{\rho}\right)
+\epsilon\left(\mu-\bar{\mu}\right)-\Psi_2
\nonumber\\
& &
+\Phi_{11}+\Lambda.
\label{r2}
\end{eqnarray}

For the case of the rescaled metric (\ref{e1}) this gives the following 
scaling behaviour with $\Omega$
\begin{eqnarray}
\begin{array}{lll}
\Phi_{00}={\cal O}(\Omega^4) \hspace{1.4cm} & \Phi_{10}={\cal O}(\Omega^3)\hspace{1.4cm}&
\Phi_{20}={\cal O}(\Omega^2)\\
\Phi_{11}={\cal O}(\Omega^2)\hspace{1.4cm}&
\Phi_{12}={\cal O}(\Omega)\hspace{1.4cm}& \Phi_{22}={\cal O}(\Omega^0)\\
\Lambda={\cal O}(\Omega^2),&&
\end{array}
\label{scal-P}
\end{eqnarray}
which determines the remaining components.
Thus the component $\Phi_{22}$ is exactly the 
one of the original space-time (\ref{pen-met}).

Therefore, Einstein's equations imply that the components of the 
energy-momentum tensor scale as
\begin{eqnarray}
\begin{array}{llll}
T_{(1)(1)}={\cal O}(\Omega^4) \hspace{1.1cm} &T_{(1)(3)}={\cal O}(\Omega^3) \hspace{1.1cm} &
T_{(1)(2)}={\cal O}(\Omega^2) \hspace{1.1cm} &T_{(3)(3)}={\cal O}(\Omega^2) \\
T_{(3)(4)}={\cal O}(\Omega^2) \hspace{1.1cm} &T_{(2)(3)}={\cal O}(\Omega) \hspace{1.1cm} &
T_{(2)(2)}={\cal O}(\Omega^0) .
\end{array}
\label{scal-T}
\end{eqnarray}
As a first example, consider a space-time containing a Maxwell field, $F_{\mu\nu}$.
Its components are given in the Newman-Penrose formalism by \cite{chandra}
\begin{eqnarray}
\phi_0&=&F_{(1)(3)}=F_{\mu\nu}l^{\mu}m^{\nu}\nonumber\\
\phi_1&=&\frac{1}{2}\left(F_{(1)(2)}+F_{(4)(3)}\right)=\frac{1}{2}F_{\mu\nu}
\left(l^{\mu}n^{\nu}+\bar{m}^{\mu}m^{\nu}\right)\nonumber\\
\phi_2&=&F_{(4)(2)}=F_{\mu\nu}\bar{m}^{\mu}n^{\nu}.
\end{eqnarray}
The non-vanishing components of the energy-momentum tensor are given by \cite{chandra}
\begin{eqnarray}
\begin{array}{lll}
T_{(1)(1)}=-2\phi_0\bar{\phi}_0\hspace{1.3cm} 
& T_{(1)(2)}+T_{(3)(4)}=-4\phi_1\bar{\phi}_1\hspace{1.3cm}&
T_{(1)(3)}=-2\phi_0\bar{\phi}_1\\
T_{(2)(2)}=-2\phi_2\bar{\phi}_2\hspace{1.3cm} &T_{(2)(3)}=-2\phi_1\bar{\phi}_2
\hspace{1.3cm} &T_{(3)(3)}=-2\phi_0\bar{\phi}_2.
\end{array}
\end{eqnarray}
Thus the components of the Maxwell field scale in the Penrose limit as follows,
\begin{eqnarray}
\phi_0={\cal O}(\Omega^2) \hspace{1.4cm}
\phi_1={\cal O}(\Omega) \hspace{1.4cm}
\phi_2={\cal O}(\Omega^0),
\end{eqnarray}
or equivalently 
\begin{eqnarray}
\phi_i={\cal O}(\Omega^{2-i}),
\end{eqnarray}
where $i=0,1,2$.
The $\phi_2$ component of the Maxwell field
is the one of the original space-time
(\ref{pen-met}).
In comparison, the peeling-off behaviour of
the components of the Maxwell tensor in  
the background of the far field of an isolated, radiating source
is given by (see for example,\cite{stew}),
\begin{eqnarray}
\phi_i={\cal O}(r^{i-3}),
\end{eqnarray}
in the limit $r\rightarrow\infty$ and
where $r$ is a radial coordinate.

As a second example, the asymptotic behaviour of the 
energy-momentum tensor describing a perfect fluid 
will be discussed.
The energy-momentum tensor is given in this case by
\begin{eqnarray}
T^{\mu\nu}=(\rho_f+p_f)u^{\mu}u^{\nu}-p_fg^{\mu\nu},
\label{T}
\end{eqnarray}
where $\rho_f$ and $p_f$ are the energy and pressure 
densities, respectively. 
Following \cite{gri} the velocity can be written 
in terms of the tetrad vectors as
\begin{eqnarray}
u^{\mu}=\frac{1}{\sqrt{2}}\left(a l^{\mu}+b n^{\mu}
\right),
\end{eqnarray}
where $a$ and $b$ are constants. For $ab=1$,
$u^{\mu}$ is a time-like vector field. This describes
the matter field in the original space-time, for
example, a perfect fluid cosmology.
In order to determine the asymptotic behaviour of 
the energy-momentum tensor, 
the constants $a$ and $b$ are  chosen such that, $a\equiv 1$
and $b\equiv\Omega^2$, resulting in 
$u_{\mu}u^{\mu}=\Omega^2\neq 1$. Furthermore, the last term in 
equation (\ref{T}) is rescaled by a factor
$\Omega^2$.
The non-vanishing components of the energy-momentum tensor
are found to be 
\begin{eqnarray}
T_{(1)(1)}&=&\frac{\Omega^4}{2}\left(\rho_f+p_f\right)
\hspace{2.7cm}
T_{(1)(2)}=\frac{\Omega^2}{2}\left(\rho_f-p_f\right)
\nonumber\\
T_{(2)(2)}&=&\frac{1}{2}\left(\rho_f+p_f\right)
\hspace{3cm}
T_{(3)(4)}=\Omega^2 p_f.
\end{eqnarray}
Thus in the limit $\Omega\rightarrow 0$ 
the fluid becomes null.
$T_{(2)(2)}$ is the only non-vanishing
component
and the scaling follows equation (\ref{scal-T}).
Furthermore, as already found in \cite{cov-pen2}
the Penrose limit of a perfect fluid 
space-time is flat  only for $p_f+\rho_f=0$, that 
is for a cosmological constant.

Einstein's equations yield the
non-vanishing
components of the Ricci tensor 
and the Ricci scalar as
\begin{eqnarray}
\Phi_{00}&=&\frac{\Omega^4}{4}\left(\rho_f+p_f\right)
\hspace{2.7cm}
\Phi_{11}=\frac{\Omega^2}{8}\left(\rho_f+p_f\right)\nonumber\\
\Phi_{22}&=&\frac{1}{4}\left(\rho_f+p_f\right)
\hspace{3cm}
\Lambda=\frac{\Omega^2}{24}\left(\rho_f-3p_f\right).
\end{eqnarray}
Thus the scaling behaviour follows equation (\ref{scal-P}).

\section{Kinematics of the null congruence}

The focussing behaviour of a congruence of 
geodesics is described by the Raychaudhuri equation,
\begin{eqnarray}
\frac{d\theta}{du}=-\frac{1}{2}\theta^2-
\sigma_{\alpha\beta}\sigma^{\alpha\beta}
+\omega_{\alpha\beta}\omega^{\alpha\beta}
-R_{\alpha\beta}k^{\alpha}k^{\beta},
\label{ra}
\end{eqnarray}
where $k^{\alpha}$
in this case  is the tangent vector
along a congruence of null geodesics (cf e.g. \cite{poi}).
This involves the expansion $\theta$, the shear
tensor $\sigma_{\alpha\beta}$ 
and the 
vorticity tensor $\omega_{\alpha\beta}$ 
of the geodesics. 
The Penrose limit of a space-time is calculated
along a null geodesic. Choosing this null 
geodesic to calculate the expansion, the shear
and the vorticity allows to determine 
further characteristics of the plane 
wave space-time obtained in the Penrose
limit, such as the occurence of caustics or
singularities.

The expansion, 
the shear and the vorticity are 
defined in terms of the null vector $k^{\alpha}$
as follows, e.g., \cite{chandra}
\begin{eqnarray}
\theta&=&\frac{1}{2}k^{\alpha}_{\; ;\alpha}\nonumber\\
\sigma^2&=&\frac{1}{2}
k^{\alpha;\beta}k_{(\alpha;\beta)}
-\frac{1}{4}\left(k^{\alpha}_{\; ;\alpha}\right)^2\nonumber\\
\omega_{\alpha\beta}\omega^{\alpha\beta}&=&
\frac{1}{2}
k^{\alpha;\beta}k_{[\alpha;\beta]},\nonumber
\end{eqnarray}
where $\sigma^2\equiv\frac{1}{2}\sigma_{\alpha\beta}\sigma^{\alpha\beta}$.

In the Penrose adapted coordinates 
(cf. metric (\ref{pen-met})) the tangent vector
along the  
congruence of null geodesics is given by the 
vector $n^{\mu}$ of the null tetrad introduced in
the previous section.
The expansion, shear and vorticity in the original
space-time with metric (\ref{pen-met}) are given by
\begin{eqnarray}
\theta&=&-\frac{1}{2}\mu_{1,u}
\label{th}\\
\sigma^2&=&
\frac{Z_{,u}\bar{Z}_{,u}}{\left(Z+\bar{Z}\right)^2}\\
\omega_{\alpha\beta}\omega^{\alpha\beta}&=&0.
\label{om}
\end{eqnarray}
There is no vorticity by construction of the coordinate
system (cf equation (\ref{pen-met})) \cite{penrose}.

Introducing the new coordinates and rescaling
the metric accordingly (cf equations 
(\ref{p1})-(\ref{e1})), shows that the
quantities (\ref{th})-(\ref{om}) are independent
of the parameter $\Omega$. Hence expansion, shear 
and vorticity are the ones characterizing the 
congruence of null geodesics in the original
space-time (\ref{pen-met}). This is consistent
with the fact that only the transverse part of 
the metric enters into these expressions.
The Raychaudhuri equation (\ref{ra})
determines the evolution of the expansion $\theta$.
However, in the case of the plane wave space-time
resulting from taking the Penrose limit, 
the kinematic quantities expansion, shear and
vorticity are already known since they are directly 
derived from the original space-time. Thus it can be 
immediately  determined if any of these
quantities becomes unbounded at some point and
thus a singularity develops.

As an example a Kasner type metric will be considered.
The metric is given by
\begin{eqnarray}
ds^2=dt^2-t^{2p_1}dx^2-t^{2p_2}dy^2-t^{2p_3}dz^2,
\label{ka}
\end{eqnarray}
where the Kasner exponents $p_i$ satisfy the relations 
$\sum_{i=1}^3p_i=1$. In vacuum there is an additional
relation, that is, $\sum_{i=1}^3p_i^2=1$.
For stiff perfect fluid cosmologies the righthandside
of this relation is equal to a constant unequal to 1.

Kasner type metrics provide a good approximation for 
the description of space-times close to a
singularity. Furthermore, for spatially homogeneous models it has been
shown that there is a class of models for which
the approach to the singularity is oscillatory and chaotic.
The dynamics can be described by a succession 
of epochs in which the Kasner indices are interchanged
\cite{bk}.

Introducing conformal time $T$, defining null
coordinates $u=T-z$, $v=T+z$ taking the Penrose limit 
according to (\ref{p1})-(\ref{p4}) and 
redefining $u$ and rescaling the remaining  
coordinates accordingly, the resulting plane wave space-time
is given by
\begin{eqnarray}
ds^2=dudv-u^{\alpha_1}dx^2-u^{\alpha_2}dy^2,
\label{K-u-met}
\end{eqnarray}
where $\alpha_1\equiv\frac{2p_1}{2-(p_1+p_2)}$ and 
$\alpha_2\equiv\frac{2p_2}{2-(p_1+p_2)}$.
Determining $\mu_1$ and $Z$ results in 
\begin{eqnarray}
\mu_1=-\frac{\alpha_1+\alpha_2}{2}\ln u\hspace{2cm}
Z=u^{\frac{\alpha_2-\alpha_1}{2}}.
\end{eqnarray}
Thus the  expansion and the shear are given by
\begin{eqnarray}
\theta&=&\frac{\alpha_1+\alpha_2}{4}\frac{1}{u}\\
\sigma^2
&=&\frac{1}{4}\left(\frac{\alpha_2-\alpha_1}{2}\right)^2
\frac{1}{u^2}.
\end{eqnarray}
In spatially homogeneous space-times the ratio of
shear over the expansion is a measure of 
anisotropy. Calculating this ratio in the 
case of the null geodesics it is found that 
\begin{eqnarray}
\frac{\sigma^2}
{\theta^2}=\left(\frac{\alpha_2-\alpha_1}{
\alpha_1+\alpha_2}\right)^2=\left(
\frac{p_2-p_1}
{p_1+p_2}\right)^2.
\end{eqnarray}
Thus the  expansion $\theta$ and the shear 
$\sigma^2$ both diverge as 
$u\rightarrow 0$. However, their ratio is constant.
Thus the amount of anisotropy measured by the ratio
$\sigma^2/\theta^2$ 
equals a constant, determined by the Kasner exponents in the 
$x$- and $y$- directions. The amount of anisotropy 
is not changed by taking the Penrose limit, that is, it is
independent of the parameter $\Omega$.

In comparison,
the anisotropy along a congruence of timelike
geodesics is determined by the constant 
ratio
\begin{eqnarray}
\frac{\sigma^2}
{\theta^2}=\frac{1}{3}\left(\sum_i p_i^2-p_1p_2
-p_1p_3-p_2p_3\right).
\end{eqnarray}

Spatially homogeneous space-times 
with form fields \cite{b6}
as well as vacuum Bianchi IX space-times show
chaotic behaviour in the approach to the initial singularity. This
behaviour is well described by a succession of Kasner epochs, each 
determined by a different set of Kasner exponents $(p_1, p_2, p_3)$.
Einstein's equations cause the effective change in the 
Kasner exponents. 
However, in the case of the Penrose limit the approach 
to the null singularity at $u=0$ is simpler than in the case of
the approach to the space-like singularity of 
the original space-time at $t=0$.
The initial set of Kasner exponents $(p_1, p_2, p_3)$ remains unchanged,
since the only Einstein equation of a plane wave is given by 
\begin{eqnarray}
2\mu_{1,uu}-\mu_{1,u}^2-4\frac{Z_{,u}\bar{Z}_{,u}}{\left(Z+\bar{Z}\right)^2}
=4\Phi_{22},
\end{eqnarray}
which in vacuum is satisfied by metrics of type 
(\ref{K-u-met}) with $\sum_i p_i=1=\sum_i p_i^2$.
The disappearance of the oscillating behaviour in the 
plane wave space-time obtained in the Penrose limit might also be understood
from the point of view of functional genericity of the metric.
General vacuum Bianchi IX space-times, for example,
are described by  8 free functions. 
In the case of  metrics of the type of equation (\ref{pen-met})
it has been argued that the most general solution of this type has
8 arbitrary functions \cite{of}.
However, taking the Penrose limit of this metric 
results in the vanishing of the 
$VV-$, the $XV-$ and $YV-$ components. Thus the 
resulting plane wave space-time has less degrees
of freedom than the general type of the original space-time.
Since chaotic behaviour in the approach to the singularity is
related to the functional genericity of the space-time
any chaotic behaviour of the original space-time is not expected to 
persist in the Penrose limit.

\section{Conclusions}
The components of the curvature tensor 
are known to show a peeling-off behaviour 
in the case of the space-time of an isolated radiating 
source or in the case of space-times
admitting two abelian Killing vectors \cite{Bondi, sp-K, bh}.
In these cases a plane wave space-time
is approached as a global limit of the 
space-time. Therefore, the peeling-off 
behaviour manifests itself as an expansion
in one of the coordinates of the 
space-time.

Here, it has been shown that in the Penrose limit  
the components of the Weyl  tensor as well 
as those of the energy-momentum tensor have 
a peeling-off property.
The Penrose limit results locally,
in the neighbourhood of a particular
null geodesic, in a plane wave space-time.
In order to obtain  the Penrose 
limit the coordinates and the metric are
rescaled by a constant parameter $\Omega$.
It is exactly this parameter which controlls
the peeling-off behaviour of the curvature
components as well as those of the 
energy-momentum tensor. 

The approach taken here is in the line
of the covariant approach to the Penrose
limit \cite{cov-pen1,cov-pen2}. Using the 
Newman-Penrose formalism it has been 
shown that the component $\Psi_4$ 
of the Weyl  tensor
of the original metric in coordinates adapted
to the neighbourhood around a null
geodesic is exactly the one that
remains the only non-vanishing component
of the Weyl tensor in the 
resulting plane wave space-time.
Similarly, $\Phi_{22}$ is the 
only non-vanishing component of the 
Ricci tensor.

The peeling-off behaviour in the 
parameter $\Omega$ might be used
to find corrections to the 
resulting plane wave space-time
also with respect to the discussion
of singularities in these type
of space-times.

The expansion, shear and vorticity of 
the congruence of null geodesic 
around which the Penrose limit is taken has
been determined. 
These expressions do not depend on the 
constant parameter $\Omega$. Their values are
determined by functions characterizing the original 
metric. By construction of the coordinate system
the congruence of null
geodesics has vanishing vorticity.

A large class of cosmological solutions 
close to the initial singularity
is well approximated by Kasner like space-times.
For this type of metrics 
it has been found that the 
anisotropy as measured by the 
ratio of shear over the expansion 
is a constant. This is the case for 
congruences of time-like as well as
null geodesics.

In the case of Bianchi IX vacuum and of spatially homogeneous
models with form fields it is known that 
the approach to the  initial (space-like)
singularity is oscillatory and chaotic.
The dynamics is described by the succession of 
epochs characterized by different Kasner exponents.
Taking the Penrose limit in one of these epochs the
corresponding Kasner indices remain unchanged.
Therefore, the approach to the appearing 
null singularity in the plane wave space-time
is not chaotic. This could also be interpreted
in terms of the functional genericity of the
plane wave space-time resulting in the Penrose limit.

\section{Acknowledgements}
This work has been supported by the programme 
``Ram\'on y Cajal'' of the M.E.C. (Spain).
Partial support by Spanish Science Ministry
Grant FPA 2002-02037 is acknowledged.

\section*{Appendix: Quantities in the Newman-Penrose formalism}

Several quantities in the Newman-Penrose
formalism are given (cf, e.g., \cite{chandra}).  The tetrad metric is given by 
\begin{eqnarray}
\eta_{(a)(b)}=\eta^{(a)(b)}=\left(
\begin{array}{cccc}
0 & 1 & 0 & 0\\
1 & 0 & 0 & 0\\
0 & 0 & 0 & -1\\
0 & 0 & -1 & 0
\end{array}
\right).
\end{eqnarray}
Tetrad indices are enclosed in brackets ( ).
They run from 1 to 4. 

For a space-time with line element (\ref{pen-met}) a null tetrad 
is provided by
\begin{eqnarray}
e_{(1)}=e^{(2)}&=&l_{\mu}=\left(
\begin{array}{cccc}
1 & \frac{\alpha}{2} &\beta_1 & \beta_2
\end{array}
\right)\nonumber\\
e_{(2)}=e^{(1)}&=&n_{\mu}=\left(
\begin{array}{cccc}
0 & 1 &0 & 0
\end{array}
\right)\nonumber\\
e_{(3)}=-e^{(4)}&=&m_{\mu}=\left(
\begin{array}{cccc}
0 & 0 &-\frac{e^{-\frac{\mu_1}{2}}}{\left(Z+\bar{Z}\right)^{\frac{1}{2}}} & 
\frac{i\bar{Z}e^{-\frac{\mu_1}{2}}}{\left(Z+\bar{Z}\right)^{\frac{1}{2}}} 
\end{array}
\right)\nonumber\\
e_{(4)}=-e^{(3)}&=&\bar{m}_{\mu}=\left(
\begin{array}{cccc}
0 & 0 &-\frac{e^{-\frac{\mu_1}{2}}}{\left(Z+\bar{Z}\right)^{\frac{1}{2}}} & 
-\frac{i Z e^{-\frac{\mu_1}{2}}}{\left(Z+\bar{Z}\right)^{\frac{1}{2}}} 
\end{array}
\right).
\end{eqnarray}
The corresponding contravariant components are given by
\begin{eqnarray}
l^{\mu}&=&\left(
\begin{array}{cccc}
-\frac{\alpha}{2} &1 & 0 &0
\end{array}
\right)\nonumber\\
n^{\mu}&=&\left(
\begin{array}{cccc}
1 & 0 &0 & 0
\end{array}
\right)\nonumber\\
m^{\mu}&=&\left(
\begin{array}{cccc}
-\frac{e^{\frac{\mu_1}{2}}}{\left(Z+\bar{Z}\right)^{\frac{1}{2}}}
\left(\beta_1\bar{Z}-i\beta_2\right) & 0&
\frac{e^{\frac{\mu_1}{2}}}{\left(Z+\bar{Z}\right)^{\frac{1}{2}}}\bar{Z} 
&-i\frac{e^{\frac{\mu_1}{2}}}{\left(Z+\bar{Z}\right)^{\frac{1}{2}}}  
\end{array}
\right)\nonumber\\
\bar{m}^{\mu}&=&\left(
\begin{array}{cccc}
-\frac{e^{\frac{\mu_1}{2}}}{\left(Z+\bar{Z}\right)^{\frac{1}{2}}}
\left(\beta_1 Z+i\beta_2\right) & 0&
\frac{e^{\frac{\mu_1}{2}}}{\left(Z+\bar{Z}\right)^{\frac{1}{2}}}Z 
&i\frac{e^{\frac{\mu_1}{2}}}{\left(Z+\bar{Z}\right)^{\frac{1}{2}}}  
\end{array}
\right).
\end{eqnarray}

The Newman-Penrose spin coefficients 
derived from the Ricci rotation coefficients, 
\newline 
$\gamma_{(a)(b)(c)}
=e_{(a)}^{\;\;\;\;\mu}e_{(b)\mu;\nu}e_{(c)}^{\;\;\;\;\nu}$,
are given by
\begin{eqnarray}
\kappa&=&\frac{e^{\frac{\mu_1}{2}}}{\left(Z+\bar{Z}\right)^{\frac{1}{2}}}
\left[\bar{Z}\left(\frac{1}{2}\beta_1\alpha_{,u}-\frac{1}{2}\alpha\beta_{1,u}
+\beta_{1,v}-\frac{1}{2}\alpha_{,x}\right)-
i\left(\frac{1}{2}\alpha_{,u}\beta_2-\frac{1}{2}\alpha\beta_{2,u}+
\beta_{2,v}-\frac{1}{2}\alpha_{,y}\right)
\right]\nonumber\\
\sigma&=&\frac{-\frac{\alpha}{2}\bar{Z}_{,u}+\bar{Z}_{,v}}{Z+\bar{Z}}
\nonumber\\
\lambda&=&-\frac{Z_{,u}}{Z+\bar{Z}}\nonumber\\
\nu&=&0\nonumber\\
\rho&=&\frac{1}{2}\left[\mu_{1,v}-\frac{\alpha}{2}\mu_{1,u}
+ie^{\mu_1}\left(\beta_1\beta_{2,u}-\beta_2\beta_{1,u}
+\beta_{1,y}-\beta_{2,x}\right)\right]\nonumber\\
\mu&=&-\frac{1}{2}\mu_{1,u}\nonumber\\
\tau&=&\frac{1}{2}\frac{e^{\frac{\mu_1}{2}}}{\sqrt{Z+\bar{Z}}}
\left(\beta_{1,u}\bar{Z}-i\beta_{2,u}\right)\nonumber\\
\pi&=&-\frac{1}{2}\frac{e^{\frac{\mu_1}{2}}}{\sqrt{Z+\bar{Z}}}
\left(\beta_{1,u}Z+i\beta_{2,u}\right)\nonumber\\
\epsilon&=&\frac{1}{4}\left[-\frac{Z_{,v}-\bar{Z}_{,v}}{Z+\bar{Z}}
+\frac{\alpha}{2}\frac{Z_{,u}-\bar{Z}_{,u}}{Z+\bar{Z}}
+ie^{\mu_1}\left[\beta_1\beta_{2,u}-\beta_2\beta_{1,u}+
\beta_{1,y}-\beta_{2,x}\right]
-\alpha_{,u}\right]\nonumber\\
\gamma&=&\frac{1}{4}\frac{\bar{Z}_{,u}-Z_{,u}}{Z+\bar{Z}}\nonumber\\
\alpha_{NP}&=&-\frac{1}{2}\left[\frac{1}{2}
\frac{e^{\frac{\mu_1}{2}}}{\sqrt{Z+\bar{Z}}}\left(\beta_{1,u}Z
+i\beta_{2,u}\right)+ie^{\mu_1}\left[
-\left(\frac{e^{-\frac{\mu_1}{2}}}{\sqrt{Z+\bar{Z}}}\right)_{,u}
\beta_2
+i\left(
\frac{Ze^{-\frac{\mu_1}{2}}}{\sqrt{Z+\bar{Z}}}\right)_{,u}\beta_1
\right.\right.\nonumber\\
& & \left.\left. +
\left(\frac{e^{-\frac{\mu_1}{2}}}{\sqrt{Z+\bar{Z}}}\right)_{,y}
-i
\left(\frac{Ze^{-\frac{\mu_1}{2}}}{\sqrt{Z+\bar{Z}}}\right)_{,x}
\right]
\right]\nonumber\\
\beta&=&-\frac{1}{2}\left[\frac{1}{2}
\frac{e^{\frac{\mu_1}{2}}}{\sqrt{Z+\bar{Z}}}\left(\beta_{1,u}\bar{Z}
-i\beta_{2,u}\right)+ie^{\mu_1}\left[
-\left(\frac{e^{-\frac{\mu_1}{2}}}{\sqrt{Z+\bar{Z}}}\right)_{,u}
\beta_2
-i\left(
\frac{\bar{Z}e^{-\frac{\mu_1}{2}}}{\sqrt{Z+\bar{Z}}}\right)_{,u}\beta_1
\right.\right.\nonumber\\
& & \left.\left. +
\left(\frac{e^{-\frac{\mu_1}{2}}}{\sqrt{Z+\bar{Z}}}\right)_{,y}
+i
\left(\frac{\bar{Z}e^{-\frac{\mu_1}{2}}}{\sqrt{Z+\bar{Z}}}\right)_{,x}
\right]
\right],
\label{sp-c}
\end{eqnarray}
where $,u$ denotes $\frac{\partial}{\partial u}$ etc.

The directional derivatives for the metric (\ref{pen-met}) are given by
\begin{eqnarray}
D&=&e_{(1)}=e^{(2)}=-\frac{\alpha}{2}\frac{\partial}{\partial u}
+\frac{\partial}{\partial v}\nonumber\\
\Delta&=&e_{(2)}=e^{(1)}=\frac{\partial}{\partial u}\nonumber\\
\delta&=&e_{(3)}=-e^{(4)}=-\frac{e^{\frac{\mu_1}{2}}}{\sqrt{Z+\bar{Z}}}
\left(\beta_1\bar{Z}-i\beta_2\right)\frac{\partial}{\partial u}
+\frac{e^{\frac{\mu_1}{2}}}{\sqrt{Z+\bar{Z}}}\bar{Z}\frac{\partial}{\partial x}
-i\frac{e^{\frac{\mu_1}{2}}}{\sqrt{Z+\bar{Z}}}\frac{\partial}{\partial y}
\nonumber\\
\bar{\delta}&=&e_{(4)}=-e^{(3)}=-\frac{e^{\frac{\mu_1}{2}}}{\sqrt{Z+\bar{Z}}}
\left(\beta_1 Z+i\beta_2\right)\frac{\partial}{\partial u}
+\frac{e^{\frac{\mu_1}{2}}}{\sqrt{Z+\bar{Z}}}Z\frac{\partial}{\partial x}
+i\frac{e^{\frac{\mu_1}{2}}}{\sqrt{Z+\bar{Z}}}\frac{\partial}{\partial y}.
\label{dd}
\end{eqnarray}

The components of the Weyl tensor 
are given by
\begin{eqnarray}
\Psi_0&=&-C_{(1)(3)(1)(3)}=-C_{\mu\nu\lambda\kappa}l^{\mu}m^{\nu}l^{\lambda}
m^{\kappa}\nonumber\\
\Psi_1&=&-C_{(1)(2)(1)(3)}=-C_{\mu\nu\lambda\kappa}l^{\mu}n^{\nu}l^{\lambda}
m^{\kappa}\nonumber\\
\Psi_2&=&-C_{(1)(3)(4)(2)}=-C_{\mu\nu\lambda\kappa}l^{\mu}m^{\nu}
\bar{m}^{\lambda}n^{\kappa}\nonumber\\
\Psi_3&=&-C_{(1)(2)(4)(2)}=-C_{\mu\nu\lambda\kappa}l^{\mu}n^{\nu}
\bar{m}^{\lambda}n^{\kappa}\nonumber\\
\Psi_4&=&-C_{(2)(4)(2)(4)}=-C_{\mu\nu\lambda\kappa}n^{\mu}\bar{m}^{\nu}
n^{\lambda}\bar{m}^{\kappa}
\end{eqnarray}

The components of the Ricci tensor are denoted as 
\begin{eqnarray}
\begin{array}{ll}
\Phi_{00}=-\frac{1}{2}R_{(1)(1)} &\hspace{1cm} \Phi_{22}=-\frac{1}{2}R_{(2)(2)}\\
\Phi_{02}=-\frac{1}{2}R_{(3)(3)} &\hspace{1cm} \Phi_{20}=-\frac{1}{2}R_{(4)(4)}\\
\Phi_{11}=-\frac{1}{4}\left(R_{(1)(2)}+R_{(3)(4)}\right) 
& \hspace{1cm}
\Phi_{01}=-\frac{1}{2}R_{(1)(3)}\\
\Lambda=\frac{1}{24}R=\frac{1}{12}\left(R_{(1)(2)}-R_{(3)(4)}\right) 
& \hspace{1cm}
\Phi_{10}=-\frac{1}{2}R_{(1)(4)}\\
\Phi_{12}=-\frac{1}{2}R_{(2)(3)} &\hspace{1cm} 
\Phi_{21}=-\frac{1}{2}R_{(2)(4)}.
\end{array}
\label{ri}
\end{eqnarray}

\end{document}